# Time-Resolved Coulomb Explosion Imaging Unveils Ultrafast Ring Opening of Furan

Enliang Wang[1,2,*], Surjendu Bhattacharyya[1], Keyu Chen[1], Kurtis Borne[1], Farzaneh Ziaee[1], Shashank Pathak[1], Huynh Van Sa Lam[1], Anbu Selvam Venkatachalam[1], Xiangjun Chen[2], Rebecca Boll[3], Till Jahnke[3,4], Artem Rudenko[1,†], Daniel Rolles[1,‡]

[1]*J. R. Macdonald Laboratory, Department of Physics, Kansas State University, Manhattan, KS, USA*

[2] *Hefei National Research Center for Physical Sciences at the Microscale and Department of Modern Physics, University of Science and Technology of China, Hefei 230026, China*

[3]*European XFEL, 22869 Schenefeld, Germany*

[4]*Max-Planck-Institut für Kernphysik, D-69117 Heidelberg, Germany*

**Abstract:** Following the changes in molecular structure throughout the entirety of a chemical reaction with atomic resolution is a long-term goal in femtochemistry. Although the development of a plethora of ultrafast technique has enabled detailed investigations of the electronic and nuclear dynamics on femtosecond time scales, direct and unambiguous imaging of the nuclear motion during a reaction is still a major challenge. Here, we apply time-resolved Coulomb explosion imaging with femtosecond near-infrared pulses to visualize the ultraviolet-induced ultrafast molecular dynamics of gas-phase furan. Widely contradicting predictions and observations for this molecule have been reported in the literature. By combining the experimental Coulomb explosion imaging data with *ab initio* molecular dynamics and Coulomb explosion simulations, we reveal the presence of a strong ultrafast ring-opening pathway upon excitation at 198 nm that occurs within 100 fs.

[*] elwang@ustc.edu.cn
[†] rudenko@phys.ksu.edu
[‡] rolles@phys.ksu.edu

Ultraviolet (UV)-induced electrocyclic ring-opening and ring-closing reactions are a prototype of pericyclic reactions, which play a crucial role in nature and in the modern chemical industry. They are also of fundamental interest for understanding non-adiabatic molecular dynamics and the role of conical intersections (CI) (*1, 2*). The UV-induced ring-opening of 1,3-cyclohexadiene, for example, is a frequently studied model system for the photobiological synthesis of vitamin D3 (*3-6*). Photocyclization reactions also have potential applications in optoelectronic devices such as optical switching (*7, 8*) and green energy storage (*9*). Therefore, ultrafast ring-opening reactions have been widely investigated by employing modern ultrafast techniques such as transient X-ray absorption (*4*), time-resolved photoelectron spectroscopy (*6, 10*), ultrafast electron diffraction (UED) (*5*), and X-ray scattering (*11*).

Here we investigate the ultrafast UV-induced ring-opening of a prototypical heterocyclic molecule, furan ($C_4H_4O$). Furan is closely related to the DNA backbone molecule tetrahydrofuran (THF, $C_4H_8O$), and both are frequently employed as the smallest model system to study the radiation damage of DNA (*28, 29*). Despite numerous theoretical and experimental investigations of the UV-induced dynamics of furan (*30-37*), a controversy remains regarding the role and relative importance of several predicted conical intersections and the resulting branching ratios of ring-puckering and ring-opening pathways. Early theoretical studies predicted a competition between the two main relaxation pathways via a series of conical intersection that either lead to ring opening or ring puckering (*34*). Along this line, subsequent experimental and theoretical work concluded that ring puckering, followed by a return to the electronic ground state of the ring-closed molecule, was by far the dominant pathway, and that only 10% of the molecules formed ring-opened isomers after passage through the puckering CI (*35, 36*). However, recent work employing transient absorption suggests that ring opening is the dominant pathway (*37*), although a different photoexcitation scheme was used in this study.

In this article, we present unambiguous evidence for a strong ring-opening pathway after 198-nm excitation by directly imaging the structural evolution of the carbon backbone during the first 500 fs of the reaction by time-resolved Coulomb explosion imaging (CEI) with a tabletop femtosecond laser. CEI was pioneered several decades ago (*12-14*) as a promising technique to determine the absolute molecular configuration of small molecules in the gas phase. The extension of this technique to time-resolved studies using femtosecond lasers and pump-probe schemes occurred as a natural next step but was mostly limited to diatomic and triatomic molecules (*15-19*). CEI relies on the idea that if all, or at least many, of the (bonding) electrons in molecule are stripped off rapidly, the structure of the remaining highly charged molecular ion is mapped to momentum space due to an explosion of the molecule by strong Coulomb repulsion. This can be triggered, e.g., by impact on an ultrathin foil (*12, 20*), charged particle impact (*21*), or through multiple ionization using X-ray free-electron laser (*22-24*) or strong-field femtosecond laser pulses (*20, 25, 26*). By employing inversion methods, it is possible to obtain three-dimensional information about every atom in the molecule. A recent review of the field was provided in (*27*).

Here, we show that time-resolved CEI can be used to directly reveal the motion of carbon atoms during an ultrafast ring-opening reaction. A schematic of the experiment is shown in Fig. 1**a**. Furan molecules were photo-excited by a UV pulse (central wavelength $\lambda$ = 198 nm, bandwidth $\Delta\lambda$ = 2 nm (FWHM), pulse duration $\Delta t$ = 105 fs (FWHM), pulse energy: 2 μJ), A strong near-infrared (NIR) pulse with an intensity of approximately $1.0 \times 10^{15}$ W/cm$^2$ ($\lambda$ = 790 nm, $\Delta\lambda$ = 52 nm (FWHM), $\Delta t$ = 28 fs) induced a subsequent Coulomb explosion of the molecule at a given time delay. The measurements were performed at a repetition rate of 10 kHz. After the Coulomb explosion, the three-dimensional momentum vectors of the ionic fragments were determined in coincidence using a COLTRIMS reaction microscope (*38, 39*).

To guide the interpretation of the experimental data, the potential energy surfaces of furan in the neutral ground state and the relevant electronically excited states were calculated by the equation-of-motion coupled-cluster method with single and double excitations (EOM-CCSD). This was followed by *ab initio* molecular dynamics (AIMD) simulations using surface hopping time-dependent density functional theory for the excited states and density functional theory for the ground state. Using the resulting trajectories, classical Coulomb explosion simulations were performed to model the effect of the time dependent molecular geometry on the CEI signal.

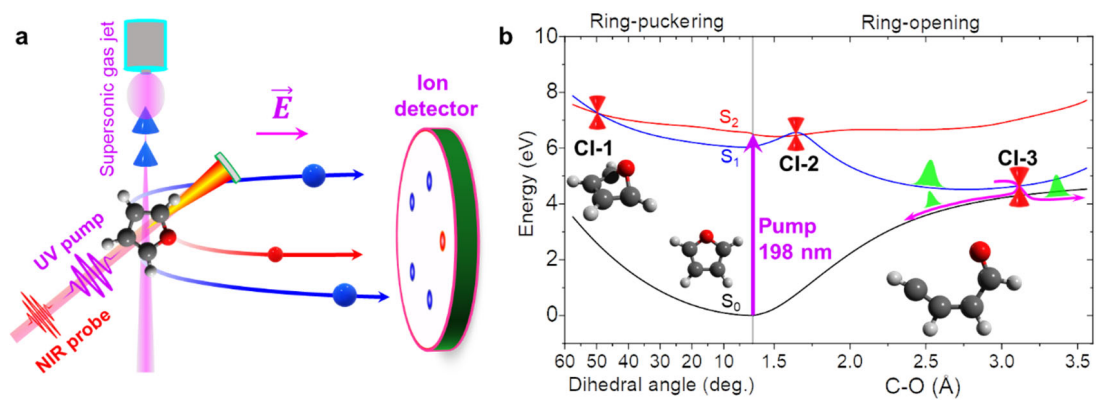

**Figure 1: Schematic of the experimental configuration (a) and the electronic structure of furan (a).** **a** A supersonic molecular beam consisting of furan molecules is irradiated with a UV pump and an NIR probe laser pulse. The resulting fragment ions are guided towards a time- and position-sensitive detector by a homogenous electric field. **b** Cuts through the potential energy surfaces, calculated at the EOM-CCSD/aug-cc-pvdz level, along the ring-opening (C-O bond length, right side) and ring-puckering (dihedral angle, left side) coordinates. For these cuts, all carbon and hydrogen atoms are kept within one plane, i.e., the molecule exhibits a transition from $C_{2v}$ to $C_s$ symmetry. For the ring-puckering coordinate, the bond lengths and the C-C-H angles are frozen and only the dihedral angle with respect to the oxygen atom is varied (i.e., the oxygen atom is moving out of plane), whereas the cut along the ring-opening coordinate is obtained by a relaxed scanning as a function of the C-O bond length. The relevant conical intersections (CI) are labeled CI-1 for the ring-puckering and CI-2 and CI-3 for the ring-opening pathway.

Starting from the molecular ground-state, the first and second excited singlet states in the vertical excitation region are due to the $S_1[^1A_2(\pi 3s)] \leftarrow S_0[^1A_1]$ and $S_2[^1B_2(\pi\pi^*)] \leftarrow S_0[^1A_1]$ transitions, respectively, with the latter corresponding to the optically bright transition (*30-32*). Upon UV (198 nm) excitation from the $S_0$ to $S_2$ state, C-O bond stretching (i.e., ring opening) and folding of the molecular dihedral angle (i.e., ring-puckering) are initiated simultaneously (see **Section 5 of the Supplementary Information (SI)** and **Supplementary Movie 1** for exemplary trajectories). Fig. 1**b** shows cuts through the potential energy surfaces (PES) along the corresponding coordinates on the right and left side of the panel, respectively. While this representation is useful to display the variation of the PES as a function of each individual degree of freedom, it is important to bear in mind that the ring-puckering and C-O bond stretching motions can happen simultaneously (*31, 34*), i.e., the molecule will not necessarily evolve exclusively along either of these two idealized paths. Nevertheless, Fig. 1**b** shows that at the EOM-CCSD/aug-cc-pvdz level of theory, the negative gradient of the PES points in the direction of the ring-opening coordinate towards two conical intersections marked as CI-2 and CI-3, whereas the reaction path towards the ring-puckering conical intersection, marked as CI-1, is slightly uphill. For the ring-puckering path, we only show the conical intersection responsible for the transition from $S_2$ to $S_1$. Further propagation along this coordinate results in a transition from $S_1$ to $S_0$ via another conical intersection, followed by a return towards the original ground-state minimum of the ring-closed molecule (*37*). In accordance with previous studies (*31*), our *ab initio* surface hopping molecular dynamics simulations (see **Section 5 of the SI**), indicate that the nonadiabatic transitions $S_2 \rightarrow S_1$ and $S_1 \rightarrow S_0$ occur after approximately 10 and 70 fs, respectively. Orbital maps for the equilibrium geometry and the conical intersection points of the $S_0$, $S_1$, and $S_2$ states are shown in **Supplementary Figure S5**. At the equilibrium geometry, the $S_2$ state has anti-bonding character, while the character of the $S_1$ state changes from 3S-Rydberg to anti-bonding along the ring-opening coordinate, which is crucial for the subsequent ring-opening reaction.

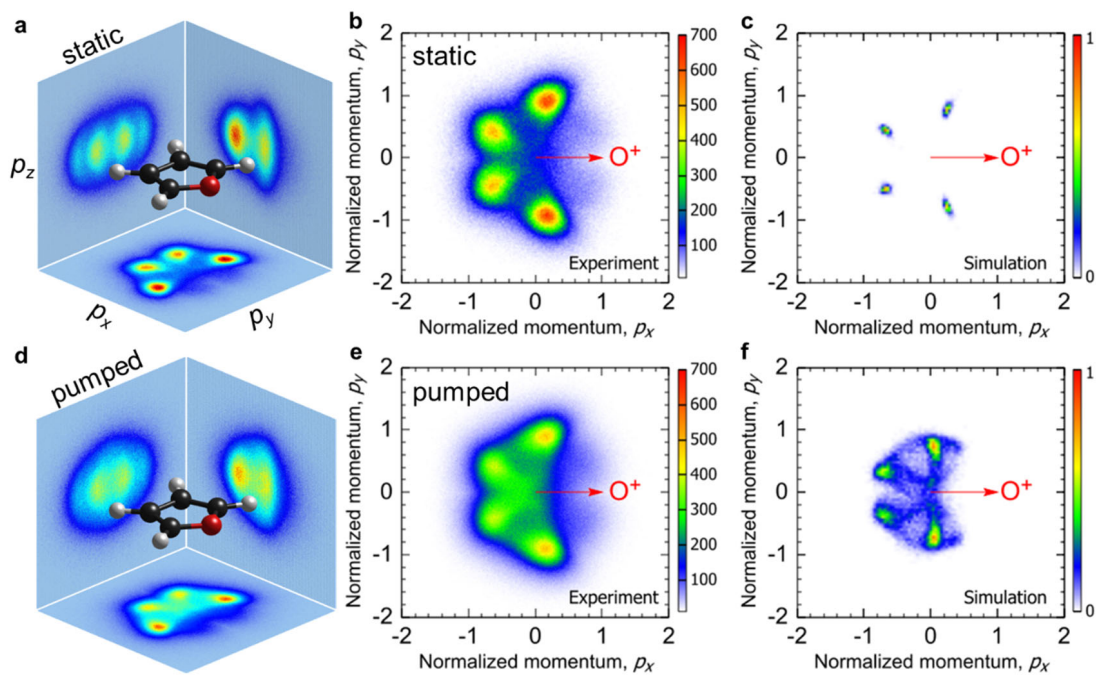

**Figure 2: Measured (a, b, d, e) and simulated (c, f) Newton plots of the oxygen and carbon momenta.** The molecular frame is defined by the $O^+$ momentum, which provides the x-axis. The x/y-plane is spanned by using the emission direction of one of the three carbon ions (randomly chosen), i.e., one of the carbon ions defines the positive direction of the y-axis. The z-axis is perpendicular to the x/y-plane. All fragment-ion momenta are normalized such that the magnitude of each $O^+$ momentum vector is equal to 1. Non-normalized Newton plots (i.e., depicting absolute momenta) are shown in Supplementary Fig. S18. **a** Projections of the normalized 3-dim. molecular-frame momenta (generated from 4-fold ($C^+$, $C^+$, $C^+$, $O^+$) coincidence events, obtained with the NIR pulses alone) onto the x/y, x/z and y/z-plane. In the x/z- and y/z-projections, the momentum of the carbon atom defining the x/y-plane is not shown since it only yields an intense line. **b** Detailed view of the x/y-projection shown in **a**. **c** Corresponding plot resulting from our theoretical modeling. **d-f** Plots corresponding to those shown in **a-c** obtained by integrating the pump-probe signals from -200 to 500 fs. **f** shows the simulated Newton plot for the opening-ring photoproducts (see text for details).

In order to experimentally investigate the molecular dynamics upon UV excitation and to determine whether ring opening is indeed a major reaction pathway or not, we have recorded time-resolved Coulomb explosion images of furan for a series of delays between the UV excitation pulse and the NIR probe pulse. As a first step, Figure 2**a** displays projections of the three-dimensional Coulomb explosion images, i.e., the molecular-frame fragment ion momentum distributions of singly charged carbon and oxygen ions resulting from the Coulomb explosion induced by the NIR pulses alone. These so-called Newton plots were constructed from ionization events where one $O^+$ and at least three $C^+$ ions were measured in coincidence. A Newton plot for the complete 5-fold fragmentation channel ($O^+ + 4\times C^+$) is shown in Supplementary Fig. **S16** and looks identical but contains less events due to the lower detection probability for 5-fold coincidences. The projection of the measured data on the molecular plane, shown in Fig. 2 **b**, is compared to the corresponding results of Coulomb explosion simulations for ground-state furan molecules with a Wigner distribution around the equilibrium geometry in Fig. 2 **c**. The positions of the maxima in the experimental and simulated plots are in excellent agreement, and the similarity of the observed patterns to the molecular structure is striking. In both, experiment and simulations, the magnitude of the momenta of the carbon ions neighboring the oxygen is slightly larger than that of the other two carbon ions. This is partly due to the C-O bond length (1.360 Å) being slightly smaller than the C-C bond lengths (1.365 Å and 1.426 Å, see Table **S1)**, and also due to the larger mass of the oxygen, which affects the momentum balance during the explosion.

To investigate how the structural changes induced by the UV excitation are reflected in the Coulomb explosion, we first inspect the Newton plots obtained by summing up all negative and positive pump-probe delays (Figs. 2**d** and **2e)**. New features that are not present in Figs. 2**a** and 2**b** are clearly discernible: the four "peaks" corresponding to the four carbon atoms in the ring are much less sharp, with significant intensity appearing in between these peaks and, in particular, filling in the center of the Newton plot near zero momentum. Both trends are also visible in our modeled results shown in Fig. 2**f**. This Newton plot obtained from the Coulomb explosion simulations includes only those ground-state AIMD trajectories that result in open-ring products. Further inspection of the simulations shows that the observed broadening of the features is a consequence of large-amplitude vibrations of the carbon atoms after UV-excitation, while the streaks towards the center of the plot are a direct and unambiguous signature of the ring opening: as the ring-open molecules adopt a more "chain-like" structure, the overall extent of the molecule increases, which decreases the momentum obtained in the Coulomb explosion. In particular, the final momenta of the atoms in the central part of the chain are lower than in the original ring-like structure, even reaching zero energy for some specific geometries (see **Section 6** of the SI for details). In addition, a slight broadening of the fragment-ion momentum distributions in the direction perpendicular to the molecular plane is observed after the UV absorption, as can be seen by comparing the corresponding projections on the two perpendicular planes in Figs. 2 **a** and **d**.

In order to directly visualize the ring-opening process as a function of pump-probe delay, we first need to establish how to construct a Newton plot for the (transient) ring-open geometries and photoproducts that no longer have the original symmetry of the furan molecule. In that case, choosing any carbon atom to define the x/y-plane, as done in Fig. 2, leads to an averaging over several possible molecular orientations. This can be avoided by taking advantage of the different kinetic energies that the different carbon atoms gain in the Coulomb explosion process depending on their position in the molecule. For furan in its ground-state equilibrium geometry, the $C_{2v}$ symmetry leads to only two distinct carbon atoms in the ring: the two atoms nearest to the oxygen are equivalent and have a higher kinetic energy after Coulomb explosion, while the two carbon atoms "opposite" to the oxygen atom, which are also equivalent, have a slightly lower kinetic energy (see **Supplementary** Fig. **S11**). Due to this symmetry, the Newton plot is uniquely defined, independent of which carbon ion is used to define the molecular plane. However, once the molecular ring opens or distorts significantly, the molecular frame depends in detail on which carbon ion is used for its definition, which leads to additional "smearing out" of the features observed in Figs 2**e**, and **f**. As detailed in **Section 6** of the **SI,** this ambiguity can be avoided to a large extend by sorting the carbon ions by their kinetic energy, as this allows to identify their original position in the ring. For example, by selecting the carbon fragment with the second highest kinetic energy, we pick either the carbon atom labeled C1 or the one marked as C2 in the inset of Fig. 3**d**, i.e., one of the C-atoms nearest to the oxygen. In the following, we use (in addition to the oxygen ion) this carbon fragment to define the molecular frame and plot this fragment in the upper half of the 2D Newton plot. This choice is mimicked in the Coulomb explosion simulations by randomly selecting C1 or C2 as a reference for the molecular plane. Figs. 3**a-d** show the Newton plots for various structures obtained from the molecular dynamics calculations employing the newly defined molecular frame. The corresponding experimental results are shown in Figs. 3**e-h** as a function of pump-probe delay (see **Section 9 of the SI** for additional plots at further delays). Both, the simulated and the measured data show a clear evolution as a function of pump-probe delay. At a delay of 50 fs, the image starts to blur due to the evolution of the wave packet on the $S_2$ and $S_1$ potential energy surfaces. As the delay increases, the maxima corresponding to C1 and C2 start bleeding into each other, and the maximum corresponding to the carbon atom at position C3 extends towards the center of the Newton plot, eventually developing a strong contribution with near-zero momentum at ≥100 fs.

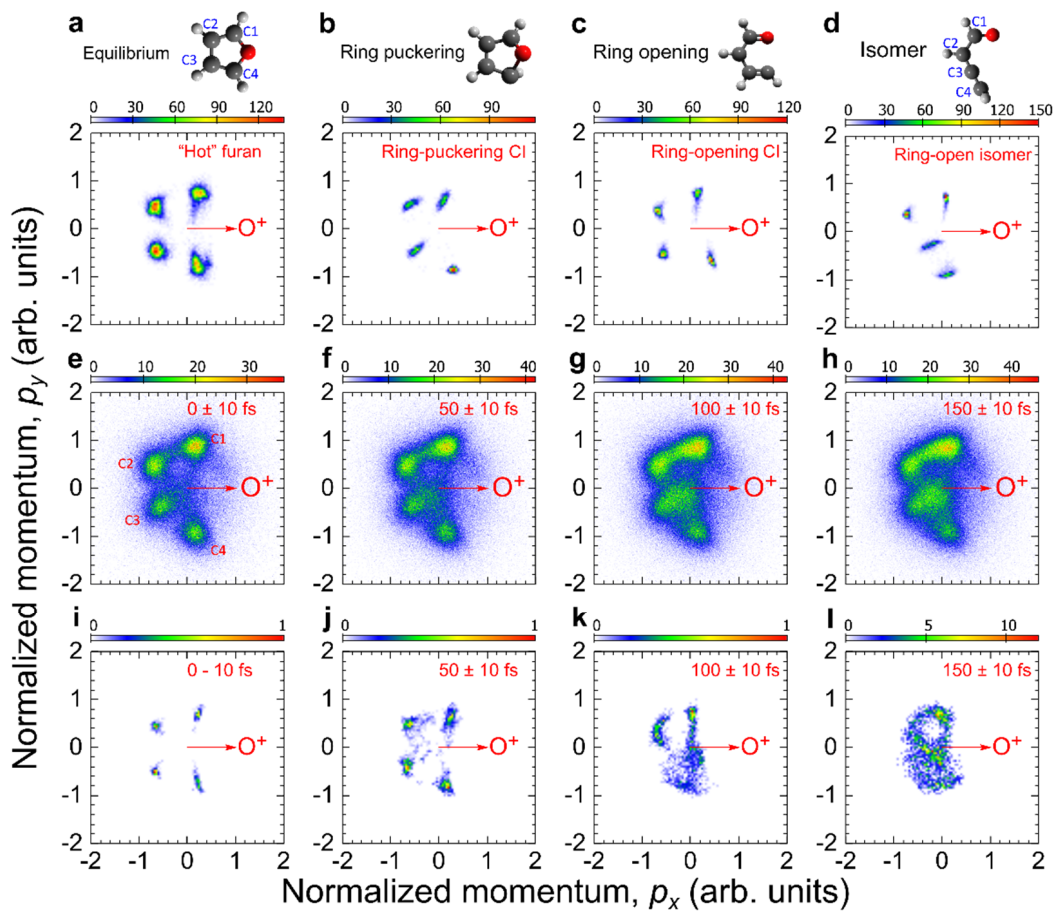

**Figure 3: Time-resolved Coulomb explosion imaging of furan.** Molecular geometry and simulated

normalized Newton plot for (**a**) "hot" ground-state furan (i.e., trajectories that have returned to the ring-closed ground state), (**b**) the ring-puckering conical intersection, (**c**) the ring-opening ($S_1 \rightarrow S_0$) conical intersection, and (**d**) for the optimized geometry of the ring-open structure, broadened by a Wigner distribution. **e-h** Time delay-snapshots of the Newton plots from the pump-probe experiment in the delay ranges indicated at the top right of each panel. **i-l** Simulated Newton plots of the ring-opening reaction, generated from the geometries obtained from the calculated trajectories on the excited (**i,j**) and the ground state (**k,l**). To allow for a direct comparison with the experiment, the start time of each ground-state trajectory has been set to 70 fs, which is the average time for the surface-hopping trajectories to reach the ground state. In all Newton plots in this figure, the molecular plane is defined by the carbon C1 or C2 (see text). Results for additional delays are shown in the Supplementary Figs. **S17 – S20**.

To help interpret these experimental observations and to link them to specific changes of the molecular structure, Coulomb explosion simulations for the geometries at the ring-puckering and ring-opening conical intersections and the equilibrium geometry of the open-ring isomer are shown in Fig. 3 **b-d**, while Coulomb explosion simulations of the ring-opening dynamics predicted by the trajectory hopping and AIMD calculations are shown in Fig. 3 **i-l**. The latter clearly reproduce some of the key observations from the experiment, such as the initial, slight blurring of the maxima, the significant smearing out of the crisp pattern at larger delay, and the characteristic shift of the C3 maximum towards the center of the plot. Especially the observation of low-momentum $C^+$ fragments is only observed in the simulations of the open-ring isomer with a nearly linear C2-C3-C4 carbon chain. Accordingly, we interpret the corresponding signal in the experiment as a clear signature of the formation and of a large yield of an open-ring product. The experimental patterns also suggest that a ring-puckering motion might be occurring in conjunction with the ring opening, but further experiments with higher temporal resolution and better statistics (which will be possible with higher-repetition-rate lasers) are needed to fully distinguish and quantify both contributions.

A more continuous and complete, albeit less visually intuitive representation of the experimental data is shown in Fig. 4, where the coincident ion yield of the $O^+ + 3C^+$ channel as a function of delay is plotted in a different molecular-frame representation. In Fig. 4**a**, the angle ϕ between the emission direction of the measured carbon ions (ranging from 0° to 360°) and the emission direction of the oxygen ion (which is located at ϕ = 0°, accordingly) is shown as a function of delay, and in Fig. 4**b**, the magnitude of the measured molecular-frame carbon momenta is inspected. At negative delays, the results are equivalent to the static (i.e. "probe-only") case. Here, rather narrow angular and momentum distributions are observed, with an angular width of the highest peak (carbon C1) of approximately 23.5° (FWHM), as obtained by a Gaussian fit to the data for t < 0 fs in Fig. 4 **a**. With increasing delay, the ion yield is strongly enhanced, and the peak width increases to 31.0° (FWHM) at 150 - 250 fs. A similar broadening for t > 0 fs is also seen in the distribution of the momentum magnitude shown in Fig. 4 **b.**

To further filter out the contamination from the unpumped and hot molecules that mainly overlap with the static image, we perform the same analysis for the carbon ion with the lowest kinetic energy, which primarily corresponds to carbon C3 in the ring-opening reaction. The resulting momentum distribution as a function of delay is shown in Fig. 4**c** along with the yields within specific momentum ranges in Fig. 4**d**. The latter shows that the relative enhancement in the yield of the lowest-energy carbon ion is stronger for smaller momenta, where the signal from ring-open products overlaps the least with that of closed-ring molecules, and that the observed dynamics occur on the time scale of the instrument response function of ≈110 fs.

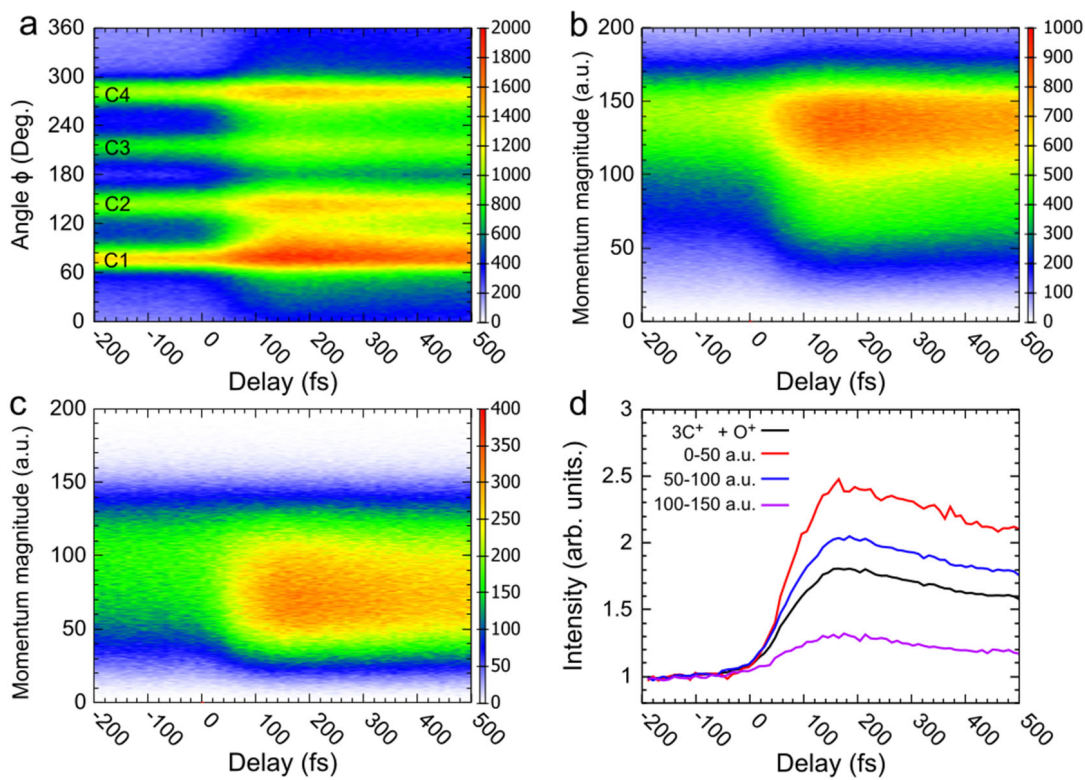

**Figure 4: Differential ion yield as a function of delay.** Coincident ion yield as a function of delay and **a** angle of the $C^+$ ion emission relative to the emission direction of the $O^+$, and **b** magnitude of the molecular-frame carbon momentum. **c** Same as **b** but restricted to the $C^+$ ion with the lowest energy. **d** Yields of carbon with the lowest energy within specific momentum magnitude ranges of **c**, compared to the total yield in the $O^+ + 3C^+$ coincidence channel (black line), as a function of delay. The data in **d** are normalized such that the average yields in the delay range from -200 to -100 fs is unity.

In summary, we have shown that time-resolved CEI, in combination with *ab initio* molecular dynamics calculations, is a powerful tool to image ultrafast changes in the molecular geometry of gas-phase molecules and directly reveal the motion of carbon atoms during an ultrafast photochemical reaction. In particular, our results provide unambiguous evidence of a strong ring-opening pathway after UV-excitation of furan at 198 nm, which was not identified in previous studies using other experimental techniques. The ring opening is found to occur within approximately 100 fs, in good agreement with surface hopping calculations. As high-repetition-rate (>10 kHz) femtosecond laser sources become more prevalent, data acquisition times for time-resolved CEI experiments will be reduce further, which should make it possible to investigate the reaction pathways in even greater detail and to image even minor reaction pathways with high fidelity.

## Methods
### Experimental setup

The experimental setup consists of a Ti:Sapphire chirped-pulse amplification laser system ($\lambda$ = 790 nm, bandwidth = 52 nm FWHM, $\Delta t$ = 28 fs, and 10 kHz repetition rate) and a coincidence ion momentum imaging spectrometer (*38-42*) The pulse duration of the 790-nm pulses was determined by a Frequency-resolved Optical Gating (FROG) measurement. The laser output was split into two beams with a 90:10 beam splitter. The lower-power branch provided the probe pulses, and the higher-power branch was used to generate the fourth harmonic probe pulses ($\lambda$ = 198 nm, bandwidth = 2 nm, FWHM) via frequency doubling and tripling of the fundamental in BBO crystals and additional sum frequency generation of the third harmonic with a fresh part of the fundamental beam in a third BBO crystal. The thicknesses of the 2$\omega$, 3$\omega$, and 4$\omega$ BBO crystals were 250, 100, and 50μm, respectively. The cross-correlation of the UV and NIR pulses was determined to be approximately 110 fs by the two-color single ionization of $C_2H_4$ gas, and the UV pulse duration was thereby derived to be approximately 105 fs. The pulse intensity of the NIR and UV were approximately $1\times10^{15}$ W/cm$^2$ and $1.7\times10^{12}$ W/cm$^2$, with the former calibrated by measuring the recoil-momentum distribution of Ne$^+$ ions following the method described in Ref. (*43*), while the latter was estimated using Gaussian beam optics for the beam parameters and focusing geometry used in this experiment.

For the molecular target, furan ($\geq$ 99.0%) was purchased from Sigma–Aldrich. Before being sent into the reaction chamber, the liquid sample was subjected to several freeze-pump-thaw cycles to minimize contamination by atmospheric gases. The gas-phase furan target was provided by a supersonic gas jet to expand into ultrahigh vacuum through a 30 μm diameter nozzle at room temperature without carrier gas. Two skimmers are used for collimating the molecular beam. At room temperature, the vapor pressure of furan is about 600 mm Hg, which is too high to perform the experiment at high laser power. To reduce the target density, a needle valve was used between the sample reservoir and the nozzle.

The created ions were extracted by a uniform electric field of 100 V/cm and detected by a multihit time- and position-sensitive delay-line detector that allows for a three-dimensional momentum vector reconstruction of all charged fragments. For the pump-probe experiment, the count rate was approximately 35 kHz, which ensures that on average, less than one molecule was fully ionized and atomized per laser pulse. The background pressure of the reaction chamber was approximately $8\times10^{-11}$ mbar, which is crucial to minimize the count rate from residual gas.

**Construction of Newton plots and identification of carbons**

The Newton plots were constructed in analogy to the recent work by Boll et al. (*24*). Molecules in the gas phase are randomly oriented in space. A molecular frame of reference was constructed by employing the measured ion momenta. The oxygen ion was selected as the reference providing the x-axis of the Newton plot. The x/y plane of the molecular frame was spanned by the emission direction of the oxygen ion and the momentum vector of one of the carbon ions. The z-axis of the molecular frame is given by the vector product of momenta of oxygen and carbon which define the molecular plane. The transformation from the laboratory frame to the molecular frame corresponds to a rotation of all measured ion momenta. Further example on how this rotation is performed is given in Section 7 of the SI.

In Fig. 2, the plots are symmetric relative to the x-axis. In this case, the molecular plane was defined by randomly choosing one of the carbon ions. In Fig. 3, the carbon fragments were distinguished by their kinetic energies, and the one with the second highest energy was used to define the upper half of the x/y-plane of the Newton plot.

The AIMD simulations showed that once the molecular ring is opened, the molecular plane is no longer well defined. The C2-C3-C4 carbon chain can rotate around the C1-C2 bond. One way to define a "molecular plane" that is comparable to the equilibrium geometry is using the oxygen, C1, and C2 ions. The Coulomb explosion simulations of the ring-opened isomer showed that C1 or C2 always have the second highest kinetic energy. Therefore, in the experiment, the carbon ion with the second highest kinetic is regarded to be the reference particle.

**Pump-probe molecular dynamics simulations**

Simulations of the molecular dynamics on the excited states were performed using the surface hopping method based on time-dependent density functional theory with Becke's three-parameter hybrid method using the Lee-Yang-Parr correlation (B3LYP) functional and 6-311G** basis set. The calculations were performed by combing the Newton-X (*44*) and Gaussian (*45*) packages. As the surface hopping calculations are time consuming, each trajectory was terminated once the electronic ground state was reached. The molecular dynamics on the ground state for both the ring-closing and ring-opening reactions were then simulated by the Atom Centered Density Matrix Propagation molecular dynamics (ADMP) method (*46-48*) with the cc-pvdz basis set using the Gaussian package (*45*). Further details regarding these simulations are given in **Section 5** of the SI. Supplementary Fig. **S7** shows the trajectories obtained by the TDDFT calculation. In agreement with the calculations by T. Fuji et al. (*45*), our simulations show that the transition from $S_1$ to $S_0$ occurs on average after approximately 70 fs. Therefore, the time-zero of the ground state simulations is shifted by +70 fs for the comparison with the experiment.

For each of the 55 simulated trajectories, the positions and velocities of every atom were extracted every 1 fs for the surface hopping and every 5 fs for the ground state simulations and used as the initial configuration for classical Coulomb explosion simulations. The Coulomb explosion simulations assume that all the atoms in the molecule are instantaneously singly ionized, i.e., Coulomb explosion of $C_4H_4O^{9+}$ is simulated. Furthermore, we assume that the interaction between the ions being purely governed by Coulomb repulsion between point charges. With these assumptions, the time-dependent positions and velocities of every atomic ion were determined by numerically solving the classical equations of motion.

**Acknowledgments**

This work was supported by the Chemical Sciences, Geosciences, and Biosciences Division, Office of Basic Energy Sciences, Office of Science, US Department of Energy, who funded EW, KC, KB, FZ, SP, HVSL, AR and DR as well as the operation of the J. R. Macdonald Laboratory through grant no. DE-FG02-86ER13491. SB was funded through grant no. DE-SC0020276 from the same funding agency and ASV through grant no. PHYS-1753324 from the National Science Foundation. EW was partially supported by the Strategic Priority Research Program of Chinese Academy of Sciences (Grant No. XDB34020000), the CAS Pioneer Hundred Talents Program, and the USTC Research Funds of the Double First-Class Initiative. The *ab initio* calculations were performed on the supercomputing system in the Supercomputing Center of the University of Science and Technology of China. We thank all the group members of the Rudenko and Rolles groups for many insightful discussions on the topic of Coulomb explosion imaging.

**References**

1. D. R. Yarkony, Diabolical conical intersections. *Rev. Mod. Phys.* **68**, 985-1013 (1996).
2. B. G. Levine, T. J. Martínez, Isomerization Through Conical Intersections. *Annu. Rev. Phys. Chem.* **58**, 613-634 (2007).
3. S. Deb, P. M. Weber, The Ultrafast Pathway of Photon-Induced Electrocyclic Ring-Opening Reactions: The Case of 1,3-Cyclohexadiene. *Annu. Rev. Phys. Chem.* **62**, 19-39 (2011).
4. A. R. Attar *et al.*, Femtosecond x-ray spectroscopy of an electrocyclic ring-opening reaction. *Science* **356**, 54-59 (2017).
5. T. J. A. Wolf *et al.*, The photochemical ring-opening of 1,3-cyclohexadiene imaged by ultrafast electron diffraction. *Nat. Chem.* **11**, 504-509 (2019).
6. J. M. Ruddock *et al.*, A deep UV trigger for ground-state ring-opening dynamics of 1,3-cyclohexadiene. *Sci. Adv.* **5**, eaax6625 (2019).
7. M. Irie, Diarylethenes for Memories and Switches. *Chem. Rev.* **100**, 1685-1716 (2000).
8. M. Irie, S. Kobatake, M. Horichi, Reversible Surface Morphology Changes of a Photochromic Diarylethene Single Crystal by Photoirradiation. *Science* **291**, 1769-1772 (2001).
9. K. Edel *et al.*, The Dewar Isomer of 1,2-Dihydro-1,2-azaborinines: Isolation, Fragmentation, and Energy Storage. *Angew. Chem. Int. Edit.* **57**, 5296-5300 (2018).
10. S. Pathak *et al.*, Tracking the ultraviolet-induced photochemistry of thiophenone during and after ultrafast ring opening. *Nat. Chem.* **12**, 795-800 (2020).
11. M. P. Minitti *et al.*, Imaging Molecular Motion: Femtosecond X-Ray Scattering of an Electrocyclic Chemical Reaction. *Phys. Rev. Lett.* **114**, 255501 (2015).

12. Z. Vager, R. Naaman, E. P. Kanter, Coulomb Explosion Imaging of Small Molecules. *Science* **244**, 426-431 (1989).
13. H. Stapelfeldt, E. Constant, H. Sakai, P. B. Corkum, Time-resolved Coulomb explosion imaging: A method to measure structure and dynamics of molecular nuclear wave packets. *Phys. Rev. A* **58**, 426-433 (1998).
14. F. Légaré *et al.*, Laser Coulomb-explosion imaging of small molecules. *Phys. Rev. A* **71**, 013415 (2005).
15. F. Légaré *et al.*, Imaging the time-dependent structure of a molecule as it undergoes dynamics. *Phys. Rev. A* **72**, 052717 (2005).
16. A. Hishikawa, A. Matsuda, M. Fushitani, E. J. Takahashi, Visualizing Recurrently Migrating Hydrogen in Acetylene Dication by Intense Ultrashort Laser Pulses. *Phys. Rev. Lett.* **99**, 258302 (2007).
17. C. B. Madsen *et al.*, A combined experimental and theoretical study on realizing and using laser controlled torsion of molecules. *J. Chem. Phys.* **130**, 234310 (2009).
18. H. Ibrahim *et al.*, Tabletop imaging of structural evolutions in chemical reactions demonstrated for the acetylene cation. *Nat. Commun.* **5**, 4422 (2014).
19. T. Endo *et al.*, Capturing roaming molecular fragments in real time. *Science* **370**, 1072-1077 (2020).
20. M. Pitzer *et al.*, Direct Determination of Absolute Molecular Stereochemistry in Gas Phase by Coulomb Explosion Imaging. *Science* **341**, 1096-1100 (2013).
21. N. Neumann *et al.*, Fragmentation Dynamics of $CO_2^{3+}$ Investigated by Multiple Electron Capture in Collisions with Slow Highly Charged Ions. *Phys. Rev. Lett.* **104**, 103201 (2010).
22. S. Zeller *et al.*, Imaging the $He_2$ quantum halo state using a free electron laser. *P. Natl. Acad. Sci. USA* **113**, 14651-14655 (2016).
23. X. Li *et al.*, Coulomb explosion imaging of small polyatomic molecules with ultrashort x-ray pulses. *Phys. Rev. Res.* **4**, 013029 (2022).
24. R. Boll *et al.*, X-ray multiphoton-induced Coulomb explosion images complex single molecules. *Nat. Phys.*,  (2022).
25. J. Voigtsberger *et al.*, Imaging the structure of the trimer systems $^4He_3$ and $^3He^4He_2$. *Nat. Commun.* **5**, 5765 (2014).
26. M. Kunitski *et al.*, Observation of the Efimov state of the helium trimer. *Science* **348**, 551-555 (2015).
27. C. A. Schouder, A. S. Chatterley, J. D. Pickering, H. Stapelfeldt, Laser-Induced Coulomb Explosion Imaging of Aligned Molecules and Molecular Dimers. *Annu. Rev. Phys. Chem.* **73**, 323-347 (2022).
28. E. Wang *et al.*, Water acting as a catalyst for electron-driven molecular break-up of tetrahydrofuran. *Nat. Commun.* **11**, 2194 (2020).
29. X. Ren *et al.*, Experimental evidence for ultrafast intermolecular relaxation processes in hydrated biomolecules. *Nat. Phys.* **14**, 1062-1066 (2018).
30. N. Gavrilov, S. Salzmann, C. M. Marian, Deactivation via ring opening: A quantum chemical study of the excited states of furan and comparison to thiophene. *Chem. Phys.* **349**, 269-277 (2008).
31. T. Fuji *et al.*, Ultrafast photodynamics of furan. *J. Chem. Phys.* **133**, 234303 (2010).
32. E. V. Gromov, A. B. Trofimov, F. Gatti, H. Köppel, Theoretical study of photoinduced ring-opening in furan. *J. Chem. Phys.* **133**, 164309 (2010).
33. E. V. Gromov, C. Lévêque, F. Gatti, I. Burghardt, H. Köppel, Ab initio quantum dynamical study of photoinduced ring opening in furan. *J. Chem. Phys.* **135**,  (2011).
34. S. Oesterling *et al.*, Substituent effects on the relaxation dynamics of furan, furfural and β-furfural:

a combined theoretical and experimental approach. *Phys. Chem. Chem. Phys.* **19**, 2025-2035 (2017).
35. S. Adachi, T. Schatteburg, A. Humeniuk, R. Mitrić, T. Suzuki, Probing ultrafast dynamics during and after passing through conical intersections. *Phys. Chem. Chem. Phys.* **21**, 13902-13905 (2019).
36. R. Spesyvtsev, T. Horio, Y.-I. Suzuki, T. Suzuki, Excited-state dynamics of furan studied by sub-20-fs time-resolved photoelectron imaging using 159-nm pulses. *J. Chem. Phys.* **143**, (2015).
37. S. Severino *et al.*, Non-Adiabatic Electronic and Vibrational Ring-Opening Dynamics resolved with Attosecond Core-Level Spectroscopy. *arXiv preprint arXiv:2209.04330*, (2022).
38. Y. Malakar *et al.*, Time-resolved imaging of bound and dissociating nuclear wave packets in strong-field ionized iodomethane. *Phys. Chem. Chem. Phys.* **21**, 14090-14102 (2019).
39. F. Ziaee *et al.*, Single- and multi-photon-induced ultraviolet excitation and photodissociation of $CH_3I$ probed by coincident ion momentum imaging. *Phys. Chem. Chem. Phys.* **25**, 9999-10010 (2023).
40. R. Dörner *et al.*, Cold Target Recoil Ion Momentum Spectroscopy: a 'momentum microscope' to view atomic collision dynamics. *Phys. Rep.* **330**, 95-192 (2000).
41. J. Ullrich *et al.*, Recoil-ion and electron momentum spectroscopy: reaction-microscopes. *Rep. Prog. Phys.* **66**, 1463 (2003).
42. C. M. Maharjan, *Momentum imaging studies of electron and ion dynamics in a strong laser field*. (Kansas State University, 2007).
43. A. Rudenko *et al.*, Resonant structures in the low-energy electron continuum for single ionization of atoms in the tunnelling regime. *J. Phys. B: At. Mol. Opt. Phys.* **37**, L407 (2004).
44. M. Barbatti *et al.*, Newton-X: a surface-hopping program for nonadiabatic molecular dynamics. *WIREs Comput. Mol. Sci.* **4**, 26-33 (2014).
45. G. W. T. M. J. Frisch, H. B. Schlegel, G. E. Scuseria, M. A. Robb, J. R. Cheeseman, G. Scalmani, V. Barone, G. A. Petersson, H. Nakatsuji, X. Li, M. Caricato, A. Marenich, J. Bloino, B. G. Janesko, R. Gomperts, B. Mennucci, H. P. Hratchian, J. V. Ortiz, A. F. Izmaylov, J. L. Sonnenberg, D. Williams-Young, F. Ding, F. Lipparini, F. Egidi, J. Goings, B. Peng, A. Petrone, T. Henderson, D. Ranasinghe, V. G. Zakrzewski, J. Gao, N. Rega, G. Zheng, W. Liang, M. Hada, M. Ehara, K. Toyota, R. Fukuda, J. Hasegawa, M. Ishida, T. Nakajima, Y. Honda, O. Kitao, H. Nakai, T. Vreven, K. Throssell, J. A. Montgomery, Jr., J. E. Peralta, F. Ogliaro, M. Bearpark, J. J. Heyd, E. Brothers, K. N. Kudin, V. N. Staroverov, T. Keith, R. Kobayashi, J. Normand, K. Raghavachari, A. Rendell, J. C. Burant, S. S. Iyengar, J. Tomasi, M. Cossi, J. M. Millam, M. Klene, C. Adamo, R. Cammi, J. W. Ochterski, R. L. Martin, K. Morokuma, O. Farkas, J. B. Foresman, and D. J. Fox, *Gaussian 09 Rev B.01*. (Gaussian, Inc., 2009.).
46. H. B. Schlegel *et al.*, Ab initio molecular dynamics: Propagating the density matrix with Gaussian orbitals. *J. Chem. Phys.* **114**, 9758-9763 (2001).
47. S. S. Iyengar *et al.*, Ab initio molecular dynamics: Propagating the density matrix with Gaussian orbitals. II. Generalizations based on mass-weighting, idempotency, energy conservation and choice of initial conditions. *J. Chem. Phys.* **115**, 10291-10302 (2001).
48. H. B. Schlegel *et al.*, Ab initio molecular dynamics: Propagating the density matrix with Gaussian orbitals. III. Comparison with Born–Oppenheimer dynamics. *J. Chem. Phys.* **117**, 8694-8704 (2002).